\documentclass[aps,prc,preprint,superscriptaddress,preprintnumbers]{revtex4-1}

\usepackage{textcomp}
\usepackage{bm}
\usepackage{amsmath}
\usepackage{mathrsfs}
\usepackage{graphicx}
\usepackage{xcolor}
\usepackage{setspace}
\usepackage{booktabs}
\usepackage{multirow}
\usepackage{amssymb}

\usepackage{CJK}
\begin{document}

\title{Solving Dirac equations on a 3D lattice with inverse Hamiltonian and spectral methods}

\author{Z.X. Ren}
 \affiliation{State Key Laboratory of Nuclear Physics and Technology,
             School of Physics, Peking University, Beijing 100871, China}%

\author{S.Q. Zhang}
\affiliation{State Key Laboratory of Nuclear Physics and Technology,
             School of Physics, Peking University, Beijing 100871, China}%

\author{J. Meng}\email{mengj@pku.edu.cn}
\affiliation{State Key Laboratory of Nuclear Physics and Technology,
             School of Physics, Peking University, Beijing 100871, China}%
\affiliation{School of Physics and Nuclear Energy Engineering,
             Beihang University, Beijing 100191, China}%
\affiliation{Department of Physics, University of Stellenbosch,
             Stellenbosch, South Africa}%

\date{\today}

\begin{abstract}
  A new method to solve the Dirac equation on a 3D lattice is proposed, in which the variational collapse problem is avoided by the inverse Hamiltonian
  method and the fermion doubling problem is avoided by performing spatial derivatives in momentum space with the help of the discrete Fourier
  transform, i.e., the spectral method. This method is demonstrated in solving the Dirac equation for a given spherical potential in 3D lattice space. In comparison with the results obtained by the shooting method,
  the differences in single particle energy are smaller than $10^{-4}$~MeV, and the densities are almost identical, which demonstrates the high accuracy of the present method.
  The results obtained by applying this method without any modification to solve the Dirac equations for an axial deformed, non-axial deformed, and octupole deformed potential are provided and discussed.
\end{abstract}

\pacs{Valid PACS appear here}
\maketitle
\section{INTRODUCTION}

The developments of new radioactive ion beam facilities and new detection techniques have largely extended our knowledge of nuclear
physics from stable nuclei to unstable nuclei far from the $\beta$-stability line, the so-called exotic nuclei. Novel and striking features have
been found in the nuclear structure of exotic nuclei, such as the halo
phenomenon~\cite{tanihata1985measurements,meng1996relativistic,meng1998giant,zhou2010neutron,Meng2015Halos} and the disappearance of traditional
magic numbers and occurrence of new ones~\cite{ozawa2000new}.
In order to describe the exotic nuclei with large space distribution, theoretical approaches should be developed in coordinate space or coordinate-equivalent space.

The density functional theory (DFT) and its covariant version (CDFT) have been proved to be effective theories for the description of exotic
nuclei~\cite{meng1996relativistic, meng1998giant,meng2016relativistic, PPNP2006,mengNPA98, dobaczewski1996mean, Bender2003Self, Pei2008Deformed}.
In comparison with its nonrelativistic counterpart, the CDFT has many attractive advantages, such as the natural
inclusion of nucleon spin freedom, new saturation property of nuclear matter~\cite{Volum16, RING1996PPNP, meng2016relativistic}, large spin-orbit splittings in
single particle energies, reproducing the isotopic shifts of Pb isotopes~\cite{sharma1993anomaly}, natural inclusion of time-odd mean field, and
explaining the pseudospin of nucleons and spin symmetries of antinucleons in
nuclei~\cite{ginocchio1997pseudospin,meng1998pseudospin,zhou2003spin,liang2015hidden}.

In most CDFT applications, the harmonic oscillator basis expansion method has been widely used, which is an very efficient approach and has achieved a great success in not only the description of the single-particle motion in nuclei \cite{liang2015hidden} but also the self-consistent description of nuclear collective modes, such as rotations \cite{Afanasjev1999PhysicsReport,Peng2008maganetic_roration,Yao2014searching,Zhao2015cranking_RHB,Zhao2015Rod-shaped,Afanasjev2016superdeform}, vibrations \cite{Niksic2002DDME1, Paar2004QRPA_spin&isospin, Vretenar2005PhysicsReport, Niksic2006BRMF_config&mix, Litvinova2006CDFT&PVC, Yao2008CPL3DAngularMP, Niksic2009BRMF_5DCH, Yao2009&3DAMP, Yao2010Config&mix_amp, LiZP20105DCH, Yao2011Config&mix_lowlying, Litvinova2011Dynamics&PVC, LiZP20115DCH, Li2013Simutaneous, Yao2014Kr74, Zhou2016molecularNe20}, and isospin excitations, by restoring the symmetries and/or considering quantum fluctuations, see also \cite{meng2016relativistic} for details.
For exotic nuclei with large spatial distribution, a large basis space is needed to get a quick convergence.
Due to the incorrect asymptotic behavior of the harmonic oscillator wave functions, this method is not appropriate for halo or giant halo nuclei \cite{dobaczewski1996mean,zhou2003spherical,PPNP2006}.
In contrast, the solution of the Dirac equation for single nucleons in coordinate space or coordinate-equivalent space is preferred.
For the spherical system, the conventional shooting method works quite well \cite{mengNPA98}, which however is rather complicated for the deformed system~\cite{Price1987Self-consistent}.
Therefore the Dirac Woods-Saxon basis expansion method was developed~\cite{zhou2003spherical} and has been widely used to solve the deformed Dirac
equation~\cite{zhou2010neutron,li2012deformed}, which, however, is highly computationally time consuming for the heavy system.

The imaginary time method (ITM)~\cite{davies1980imaginary} is a powerful approach for the self-consistent mean-field calculations in a
three-dimensional (3D) coordinate space.
The ITM has been successfully employed in nonrelativistic self-consistent mean-field
calculations~\cite{bonche2005solution,Maruhn2014CPC}. For a long time, there exist doubts about the access of the ITM to the Dirac equation due to the Dirac sea, i.e., the relativistic ground state within the Fermi sea is a saddle point rather than a minimum. This is the so-called variational collapse
problem~\cite{zhang2009first,zhang2009solving,ZhangIJMPE2010,hagino2010iterative,tanimura20153d}. To avoid the variational collapse, Zhang \textit{et al }~\cite{zhang2009first,ZhangIJMPE2010} applied the ITM to the Schr\"{o}dinger-equivalent form of the Dirac equation in the spherical case. The same
method is used to solve the Dirac equation with a nonlocal potential in Refs.~\cite{zhang2009first,zhang2009solving}. Based on the idea of Hill and
Krauthauser~\cite{hill1994solution}, Hagino and Tanimura proposed the inverse Hamiltonian method (IHM) to avoid variational
collapse~\cite{hagino2010iterative}. This method solves the Dirac equation directly and the Dirac spinor is obtained
simultaneously.

Meanwhile when the IHM method is applied to lattice space in numerical calculations, another challenge appears, i.e.,  fermion doubling problem~\cite{salomonson1989relativistic,tanimura20153d} due to the replacement of the
derivative by the finite difference method~\cite{salomonson1989relativistic,tanimura20153d}.
This problem appears also in lattice quantum chromodynamics (QCD)~\cite{wilson1977new,kogut1983lattice}, which has been solved by Wilson's fermion method~\cite{wilson1977new,kogut1983lattice}. In Ref.~\cite{tanimura20153d}, Tanimura, Hagino, and Liang followed the same idea and realized the relativistic calculations on 3D lattice by introducing high-order Wilson term. However, the high-order Wilson term modified the original Dirac Hamiltonian and the single particle energies and wave functions need be corrected. Although the corrections can be done with the perturbation theory, numerically it is much more involved.
Another problem is that the high-order Wilson term introduces artificial symmetry breaking to the system~\cite{tanimura20153d}.

In this paper, we propose a new recipe for the imaginary time method to solve the Dirac equation in 3D lattice space, where the variational collapse
problem is avoided by the IHM, and the Fermion doubling problem is avoided by performing the spatial derivatives of the Dirac equation in momentum space
with the help of discrete Fourier transform, the so-called spectral method~\cite{Shen2011Spectral}.

This method is demonstrated by solving the Dirac equation for a given spherical potential in 3D lattice space and  comparing with the results obtained by the shooting method.
By extending this method to solve the Dirac equations for an axial deformed, non-axial deformed, and octupole deformed potential, the corresponding single particle energy levels are obtained.
The corresponding quantum numbers of these energy levels are obtained respectively by projection.

The paper is organized as follows, the variational collapse and the Fermion doubling problems will be briefly
introduced in Sec. \ref{numerical_method} together with the inversion Hamiltonian method and the spectral method. In Sec. \ref{Numericaldetails} the parameters for Woods-Saxon type potentials and the numerical details are presented. Sec. \ref{results&discussion} is devoted to results and discussions. Summary and perspectives are given in Sec. \ref{summary}.

\section{THEORETICAL FRAMEWORK}\label{numerical_method}
   \subsection{VARIATIONAL COLLAPSE AND INVERSE HAMILTONIAN METHOD}
       \subsubsection{IMAGINARY TIME METHOD}

The ITM is an iterative method for mean-field problem. The idea of ITM is to replace time with an imaginary number, and the evolution of the wave function reads~\cite{davies1980imaginary},
\begin{equation}
   \textrm{e}^{-{\rm i}\hat{h}t}|\psi_0\rangle\xrightarrow{t\rightarrow-{\rm i}\tau}\textrm{e}^{-\hat{h}\tau}|\psi_0\rangle,
\end{equation}
where $|\psi_0\rangle$ is an initial wave function and $\hat{h}$ is the Hamiltonian.

With the eigenstates $\{\phi_k\}$ of the Hamiltonian $\hat{h}$ corresponding to the eigenenergies $\{\varepsilon_k\}$, the evolution of the wave function
$|\psi(\tau)\rangle=\textrm{e}^{-\hat{h}\tau}|\psi_0\rangle$ can be written as,
\begin{equation}\label{Eq_imaginary}
  |\psi(\tau)\rangle=\textrm{e}^{-\hat{h}\tau}|\psi_0\rangle=\sum_k\textrm{e}^{-\varepsilon_k\tau}|\phi_k\rangle\langle\phi_k|\psi_0\rangle,
\end{equation}
where $\varepsilon_1\leq\varepsilon_2\leq\cdots$.
For $\tau\rightarrow\infty$, $|\psi(\tau)\rangle$ will approach the ground state
wave function of $\hat{h}$ as long as $\langle\phi_1|\psi_0\rangle\neq0$.

In practice, the imaginary time $\tau$ is discrete with the interval $\Delta \tau$, i.e., $\tau = N \Delta \tau$.
The wave function at $\tau=(n+1)\Delta \tau$ is obtained from the wave function at $\tau=n\Delta \tau$ by expanding the exponential evolution operator $\textrm{e}^{-\Delta \tau\hat{h}}$ to the linear order of $\Delta \tau$,
\begin{equation}\label{Eq_iterative}
   |\psi^{(n+1)}\rangle\propto\left(1-\Delta \tau\hat{h}\right)|\psi^{(n)}\rangle.
\end{equation}
Since this evolution is not unitary, the wave function should be normalized at every step.

In order to find excited states, one can start with a set of initial wave functions and orthonormalize them during the evolution by the Gram-Schmidt method. This method has been successfully
employed in the 3D coordinate-space calculations for nonrelativistic systems \cite{bonche2005solution,Maruhn2014CPC}.

   \subsubsection{VARIATIONAL COLLAPSE}

For the static Dirac equation,
\begin{equation}\label{Dirac_equation}
  \{-\textrm{i}\bm{\alpha\cdot\nabla}+V(\bm{r})+\beta[m+S(\bm{r})]-m\}\psi(\bm{r})=\varepsilon\psi(\bm{r}),
\end{equation}
with $\bm{\alpha}$ and $\beta$ the Dirac matrix, $V(\bm{r})$ the vector  potential, $S(\bm{r})$ the scalar potential, and $\psi(\bm{r})$ the Dirac spinor, its eigenenergy spectrum extends from the continuum in the Dirac sea to the continuum in the Fermi sea.
Because of the existence of the Dirac sea, the evolution in Eq. \eqref{Eq_imaginary} inevitably dives into the Dirac sea (negative energy states) as $\tau\rightarrow\infty$, which is the so-called variational collapse problem~\cite{ZhangIJMPE2010}.

   \subsubsection{INVERSE HAMILTONIAN METHOD}

To avoid the variational collapse, Hagino and Tanimura proposed the inverse Hamiltonian method~\cite{hagino2010iterative} to find the wave function of the
Dirac Hamiltonian $\hat{h}$ by,
\begin{equation}\label{Eq_inverseH}
  \lim_{\tau\rightarrow\infty}\textrm{e}^{\tau/(\hat{h}-W)}|\psi_0\rangle,
\end{equation}
where $W$ is an auxiliary parameter introduced to locate the interested eigenstate.

With a given $W$, the spectrum of $\hat{h}$ can be labeled as
\begin{equation}
  \cdots\leq\varepsilon_{-2}\leq\varepsilon_{-1}<W<\varepsilon_1\leq\varepsilon_2\leq\cdots,
\end{equation}
where $\cdots, \varepsilon_{-2},\varepsilon_{-1}$ and $\varepsilon_{1}, \varepsilon_{2}, \cdots$ are the eigenenergies of the Dirac Hamiltonian $\hat{h}$. Accordingly, the spectrum of $1/(\hat{h}-W)$ reads,
\begin{equation}
   \frac{1}{\varepsilon_{-1}-W}\leq\frac{1}{\varepsilon_{-2}-W}\leq\cdots \leq\frac{1}{\varepsilon_{2}-W}\leq\frac{1}{\varepsilon_{1}-W}.
\end{equation}
The evolution of the wave function in Eq. \eqref{Eq_inverseH} will lead to the eigen wave function $|\phi_1\rangle$ corresponding to the eigenvalue $\varepsilon_{1}$,
\begin{align}\label{Eq_iterativeIHM}
  &\lim_{\tau\rightarrow\infty}\textrm{e}^{\tau/(\hat{h}-W)}|\psi_0\rangle\nonumber\\
  &~~=\lim_{\tau\rightarrow\infty}\sum_k\textrm{e}^{\tau/(\varepsilon_k-W)}
  |\phi_k\rangle\langle\phi_k|\psi_0\rangle\propto|\phi_1\rangle,
\end{align}
as long as $\langle\phi_1|\psi_0\rangle\neq0$.

In practice, the imaginary time evolution in Eq. \eqref{Eq_inverseH} is performed iteratively,
\begin{equation}
   |\psi^{(n+1)}\rangle \propto \left(1+\frac{\Delta \tau}{\hat{h}-W}\right)|\psi^{(n)}\rangle.
   \label{Eq_inversWF}
\end{equation}
The wave function also should be normalized at every step.
The inverse of the Hamiltonian in Eq. \eqref{Eq_inversWF},
$\displaystyle \frac{\Delta \tau} {\hat{h}-W} | \psi^{(n)}\rangle$,
can be solved iteratively by the conjugate residual method~\cite{saad2003iterative}.

To find excited states, with a set of initial wave functions there are two options for choosing $W$.
One can take a fixed $W$, then evolve the set of wave functions and orthonormalize them during the evolution by the Gram-Schmidt method.
Alternatively, one can take the set of $W_i$ for each eigenstate $i$ to evolve the whole set of wave functions. The details can be found in Sec.~\ref{Numericaldetails}, where an efficient method for choosing $W_i$ is suggested to achieve  a fast convergence.

   \subsection{FERMION DOUBLING PROBLEM AND SPECTRA METHOD}
      \subsubsection{FERMION DOUBLING PROBLEM}

For a Dirac equation on 3D lattice, there exists a so-called Fermion doubling problem due to
the replacement of the first derivatives in the Dirac equation \eqref{Dirac_equation} by the finite difference
method \cite{salomonson1989relativistic,tanimura20153d}. Taking the one-dimensional Dirac equation as an example,
\begin{equation}\label{Eq_1dDirac}
  (-\textrm{i}\alpha\partial_x+\beta m)\psi(x)=\varepsilon\psi(x),
\end{equation}
its solution has the form
\begin{equation}
  \psi(x)=\tilde{\psi}(k)\exp(\textrm{i}kx).
\end{equation}
If one approximates the derivative $\partial_x$ in Eq. \eqref{Eq_1dDirac} with a three-point differential formula with the mesh size $d$, the Dirac
equation \eqref{Eq_1dDirac} becomes,
\begin{equation}\label{Eq_3pointDirac}
  \left[\frac{1}{d}\alpha\sin(kd)+\beta m\right]\tilde{\psi}(k)=\varepsilon\tilde{\psi}(k).
\end{equation}
The dispersion relation obtained from Eq.~\eqref{Eq_3pointDirac} reads,
\begin{equation}\label{Eq_3pointDispersion}
  \varepsilon^2=\frac{1}{d^2}\sin^2(kd)+m^2,
\end{equation}
which differs from the exact one,
\begin{equation}\label{Eq_exactDispersion}
  \varepsilon^2=k^2+m^2.
\end{equation}
For the dispersion relation \eqref{Eq_3pointDispersion} obtained with the three-point differential formula, there are two momenta corresponding to one energy in the momentum interval $[0,d/\pi]$. The lower momentum corresponds to the physical solution, while the higher momentum corresponds to a spurious solution. As illustrated in Ref.~\cite{tanimura20153d}, this problem persists even with the more accurate finite differential formula.
Similar spurious solution problem in radial Dirac equations are also demonstrated in Ref.~\cite{zhao2016spherical}.

   \subsubsection{SPECTRAL METHOD}

To avoid the Fermion doubling problem, the derivative in Eq.~\eqref{Eq_1dDirac} can be performed in momentum space,
\begin{equation}
    \left[\alpha k+\beta m\right]\tilde{\psi}(k)=\varepsilon\tilde{\psi}(k),
\end{equation}
which yields the exact dispersion relation; i.e., the fermion doubling problem is avoided naturally.
This is the so-called spectral method, i.e., to perform spatial derivatives in momentum space. In the following, this method is illustrated in a 1D case and it is straightforward to generalize this method to the 3D case.

We assume that there are even $n_x$ discrete grid points $x_\nu$ in coordinate space distributing symmetric with
the origin point,
\begin{equation}
     x_\nu=\left(-\frac{n_x-1}{2}+\nu-1\right)dx,~~\nu=1,...,n_x,
\end{equation}
same number of grid points $k_\mu$ in momentum space,
\begin{equation}
  k_\mu=
    \begin{cases}
      (\mu-1)dk,&\mu=1,...,n_x/2,\\
      (\mu-n_x-1)dk,&\mu=n_x/2+1,...,n_x,
    \end{cases}
\end{equation}
 and the steps in coordinate space $dx$ and in momentum space $dk$ are related by,
\begin{equation}
    dk=\frac{2\pi}{n_x\cdot dx}.
\end{equation}

The function in coordinate space $f(x_\nu)$ and the function in momentum space $\tilde{f}(k_\mu)$ are connected by
the discrete Fourier transform,
\begin{subequations}
  \begin{align}
    &\tilde{f}(k_\mu)=\sum_{\nu=1}^{n_x}\exp(-\textrm{i} k_\mu x_\nu)f(x_\nu),\label{Eq_DFT}\\
    &f(x_\nu)=\frac{1}{n_x}\sum_{\mu=1}^{n_x}\exp(\textrm{i} k_\mu x_\nu)\tilde{f}(k_\mu).\label{Eq_inversDFT}
  \end{align}
\end{subequations}
From Eq.\eqref{Eq_inversDFT}, the $m$-th order derivative of $f(x_\nu)$ can be found as,
\begin{equation}
  \begin{split}
     f^{(m)}(x_\nu)&=\frac{1}{n_x}\sum_{\mu=1}^{n_x}\exp(\textrm{i} k_\mu x_\nu)(\textrm{i} k_\mu)^m\tilde{f}(k_\mu)\\
                   &=\frac{1}{n_x}\sum_{\mu=1}^{n_x}\exp(\textrm{i} k_\mu x_\nu) \tilde{f}^{(m)}(k_\mu).
  \end{split}
\end{equation}
Here $\tilde{f}^{(m)}(k_\mu)$ corresponds to the Fourier transform of the $m$-th order derivative of $f(x_\nu)$,
\begin{equation}\label{Eq_f&fm}
  \tilde{f}^{(m)}(k_\mu)=(\textrm{i} k_\mu)^m\tilde{f}(k_\mu).
\end{equation}

In summary, the procedures to perform derivatives in coordinate space are as follows:
(1) calculate $\tilde{f}(k_\mu)$ from $f(x_\nu)$ by the discrete Fourier transform in Eq.~\eqref{Eq_DFT};
(2) calculate $\tilde{f}^{(m)}(k_\mu)$ by Eq. \eqref{Eq_f&fm};
(3) calculate the $m$-th order derivative $\tilde{f}^{(m)}(x_\nu)$ from $\tilde{f}^{(m)}(k_\mu)$ by
the inverse discrete Fourier transform as in Eq. \eqref{Eq_inversDFT}.

The spectral method has the advantage to perform the spatial derivatives with a good accuracy. The information of all grids is used in calculating the spatial derivative of any grid. Different from the finite differential method,
all grids are treated on the same footing and the grids near the boundaries do not need special numerical techniques.

\section{NUMERICAL DETAILS}\label{Numericaldetails}

In the following, we will solve the Dirac equation on 3D lattice in which the variational collapse problem is avoided by the inverse Hamiltonian method, and the fermion doubling problem is avoided by performing spatial derivatives in momentum space with the help of the discrete Fourier transform, i.e., spectral method.

The vector potential $V(\bm{r})$ and the scalar potential $S(\bm{r})$ in Eq.~\eqref{Dirac_equation} are Woods-Saxon type potentials satisfying,
\begin{equation}\label{Eq_potential}
  \begin{split}
    &V(\bm{r})+S(\bm{r})=\frac{V_0}{1+\exp[(r-R_0F(\Omega))/a]},\\
    &V(\bm{r})-S(\bm{r})=\frac{-\lambda V_0}{1+\exp[(r-R_{ls}F(\Omega))/a_{ls}]},
  \end{split}
\end{equation}
where $F(\Omega)$ is a function of $\Omega=(\theta, \varphi)$ with potential deformation parameters $\beta_{20}$, $\beta_{22}$ and $\beta_{30}$,
\begin{equation}\label{Eq_deformed}
   F(\Omega)=1+\beta_{20}Y_{20}(\Omega)+\beta_{22}[Y_{22}(\Omega)+Y_{2(-2)}(\Omega)]+\beta_{30}Y_{30}(\Omega).
\end{equation}
The deformation parameters $\beta_{20}$ and $\beta_{22}$ in Eq.~\eqref{Eq_deformed} are related to Hill-Wheeler coordinates $\beta$ and $\gamma$~\cite{Hill1953PhysicalReview,ring2004nuclear} by
\begin{equation}\label{Eq_HWcoordinates}
  \begin{cases}
    \beta_{20}=\beta\cos\gamma,\\
    \beta_{22}=\frac{1}{\sqrt{2}}\beta\sin\gamma.
  \end{cases}
\end{equation}
The adopted Woods-Saxon potential parameters in Eq. \eqref{Eq_potential} are listed in Table \ref{Tab_parameter}, which correspond to the neutron potential in
$^{48}$Ca~\cite{koepf1991WoodsSaxon}.
\begin{table}[h]
\caption{\label{Tab_parameter}The parameters in the Woods-Saxon type potential Eq. \eqref{Eq_potential} adopted in the present 3D lattice calculations.}
\begin{ruledtabular}
\begin{tabular}{cccccc}
$V_0$ [MeV]     &$R_0$ [fm]     &$a$ [fm]    &$\lambda$     &$R_{ls}$ [fm]      &$a_{ls}$ [fm]\\
\hline
-65.796        &4.482         &0.615      &11.118        &4.159             &0.648
\end{tabular}
\end{ruledtabular}
\end{table}

In the calculations, the box sizes $L=23$~fm and step sizes $d=1$~fm are respectively chosen along $x$, $y$ and $z$ axes if not otherwise specified. The imaginary time step size $\Delta T$ is taken 100~MeV.

For the $i$-th level, the upper component of the initial wave function is generated from a nonrelativistic harmonic oscillator state  and the corresponding lower component is taken the same as the upper one.
The energy shift $W_i$ is taken as
\begin{equation}
  W_i=\varepsilon_i - \Delta W_i,
\end{equation}
where $\varepsilon_i$ is the expectation value of the Dirac Hamiltonian for the $i$-th level.
The choice of $\Delta W_i$ is as follows: $\Delta W_1=6$ MeV and for $i > 1$,
\begin{equation}
 \begin{split}
   &~~\Delta W_i =
     \begin{cases}
       \varepsilon_i-\varepsilon_{i-1},&\varepsilon_i-\varepsilon_{i-1} > \Delta W_1 \\
       \Delta W_{i-1},&\varepsilon_i-\varepsilon_{i-1}\leqslant \Delta W_1
     \end{cases}
  \end{split}
\end{equation}

The convergence in the evolution of the wave functions for our interested states is determined by $\sqrt{\langle \hat{h}^2 \rangle_i - \langle\hat{h}\rangle_i^2}$ smaller than the required accuracy
$\delta_i = 10^{-4}$~MeV if not otherwise specified.

To speed up the convergence, the Dirac Hamiltonian is diagonalized within the space of the evolution wave functions every 10 iterations, and the eigenfunctions thus obtained are taken as initial wave functions for future iteration. A similar technique is also used in Ref.~\cite{Maruhn2014CPC}.

\section{RESULTS AND DISCUSSION}\label{results&discussion}
    \subsection{SPHERICAL POTENTIAL}

In this section, the Dirac equation with a given potential is solved in 3D lattice space by the new method (denoted as \textit{3D lattice}). First we examine the convergence feature of the present 3D lattice calculation for a spherical potential in Eq. \eqref{Eq_potential}.
The results will be compared with those obtained by the shooting method (denoted as \textit{shooting})~\cite{mengNPA98} with a box size $R=20$~fm and a step size $dr=0.01$~fm.

\begin{figure}
  \centering
  \includegraphics[width=0.45\textwidth]{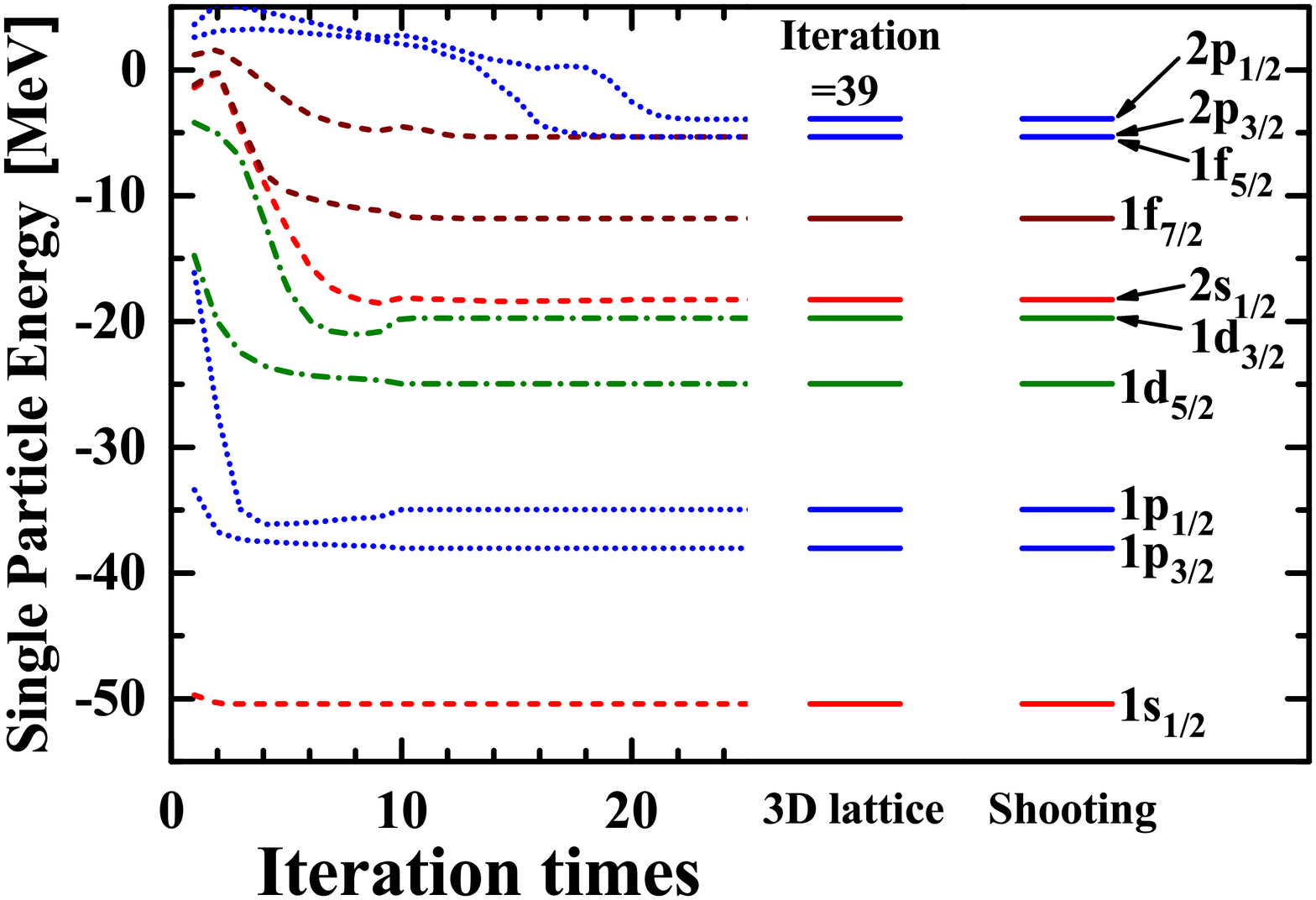}\\
  \caption{Evolution of single particle energies in the spherical Woods-Saxon potential in Eq. \eqref{Eq_potential} as a function of
  iteration times. Convergence is achieved after the 39th iteration where the energy dispersions of all bound single particle levels are
  smaller than $10^{-4}$~MeV. As a comparison, the results obtained by the shooting method are also given.
  }\label{Fig_iteration}
\end{figure}

With the potential parameters in Table \ref{Tab_parameter}, the evolution of single particle energies as a function of iteration times is shown in Fig. \ref{Fig_iteration}. There are in total of 40 bound single particle states in the 3D lattice calculation and some of them are degenerate in energy due to the spherical symmetry.
For clarity, only one energy level of the degenerate ones is shown to illustrate the evolution of single particle energies.
The single particle energies obtained by the shooting method are also shown for comparison. It can be seen that the deeper levels converge more quickly. After the 39th iteration, the accuracy of energy for all bound levels is smaller than $10^{-4}$~MeV. A distinct feature is observed at the 10th iteration where the convergence of 1p$_{1/2}$, 1d$_{3/2}$, and 2s$_{1/2}$ states is speeded up due to the diagonalization of the Hamiltonian within the space of the evolution wave functions. In fact, it will cost tens of thousands of iteration steps to reach the convergence tolerance without this diagonalization procedure.

\begin{figure}[h]
  \centering
  \includegraphics[width=0.45\textwidth]{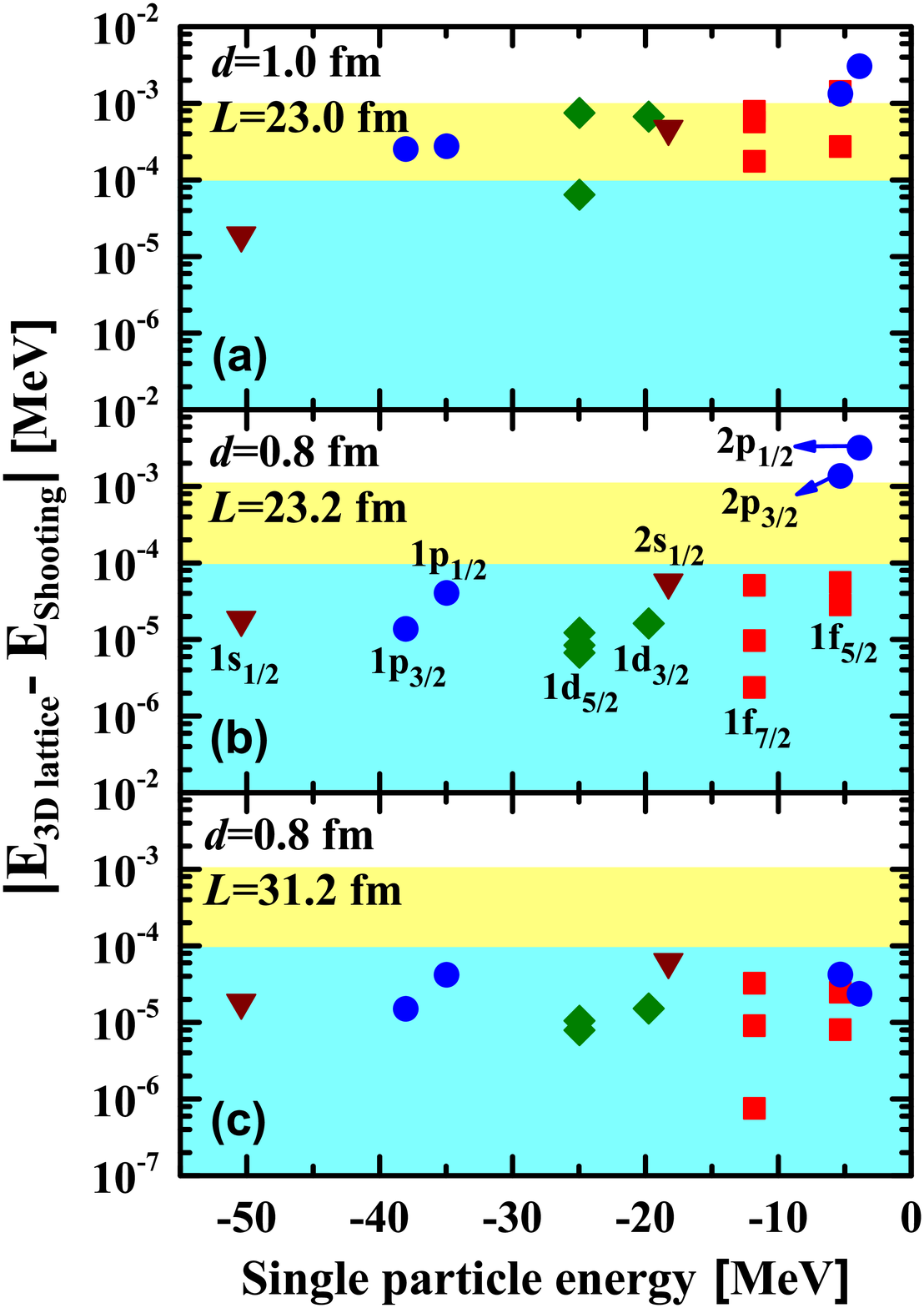}\\
  \caption{Absolute deviations of single particle energies between the 3D lattice calculation and the shooting method as a function of single particle energy for (a) step sizes $d=1.0$ fm and box sizes $L=23.0$ fm, (b) $d=0.8$ fm and $L=23.2$ fm, and (c) $d=0.8$ fm and $L=31.2$ fm.
  The spherical quantum numbers are listed in (b).
  }\label{Fig_Energydifference}
\end{figure}

In Fig. \ref{Fig_Energydifference}, the absolute deviations of single particle energies between the 3D lattice calculation and the shooting method are given as a function of single particle
energy for different step sizes $d$ and box sizes $L$.
In Fig. \ref{Fig_Energydifference} (a), for $d=1.0$ fm and $L=23.0$ fm, the absolute deviations of single particle energies are smaller than $10^{-3}$ MeV, except the weakly
bound states 1f$_{5/2}$, 2p$_{3/2}$, and 2p$_{1/2}$.
In Fig. \ref{Fig_Energydifference} (b), for $d=0.8$ fm and $L=23.2$ fm, the absolute deviations of single particle states are less than $10^{-4}$ MeV, except 2p$_{3/2}$ and 2p$_{1/2}$.
And in Fig. \ref{Fig_Energydifference} (c), for $d=0.8$ fm and $L=31.2$ fm, all absolute deviations including 2p$_{3/2}$ and 2p$_{1/2}$ are smaller than $10^{-4}$MeV.

These results indicate that smaller step size can definitely improve the accuracy but not for the weakly bound states with low orbital angular momentum. By choosing suitable step and box sizes, accurate descriptions for all the bound states including the weakly bound states 2p$_{3/2}$ and 2p$_{1/2}$ can be achieved in the 3D lattice calculations.

\begin{figure}[h]
  \centering
  \includegraphics[width=0.45\textwidth]{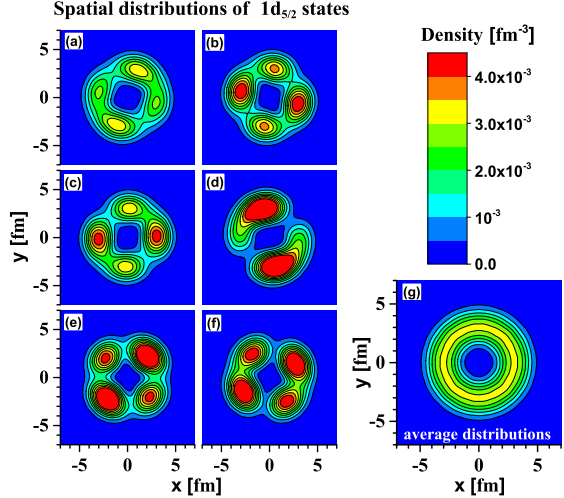}\\
  \caption{Spatial distributions of the states corresponding to 1d$_{5/2}$ in $z=0$ plane in the 3D lattice calculation.
  Figures (a)-(f) are the density
  distributions of the states in 1d$_{5/2}$, and (g) is their average spatial distributions.}
  \label{Fig_densityplane}
\end{figure}

It is interesting to investigate the spatial distributions of states and examine their agreements with the results obtained by the shooting method.
In Fig.~\ref{Fig_densityplane}, as examples, the distributions of the states corresponding to 1d$_{5/2}$ in $z=0$ plane are illustrated.
The states corresponding to 1d$_{5/2}$ are six degenerate single-particle states in the 3D lattice calculations.
Their spatial distributions are respectively shown in Figs.~\ref{Fig_densityplane} (a)-(f), and Fig.~\ref{Fig_densityplane} (g) exhibits their average in the $z=0$ plane.
As there is no symmetry restriction in the 3D lattice calculations, the six states are randomly oriented in space.
However, their average spatial distribution does show the spherical symmetry as shown in Fig.~\ref{Fig_densityplane} (g), which is consistent with the given spherical potential.

\begin{figure}[h]
  \centering
  \includegraphics[width=0.45\textwidth]{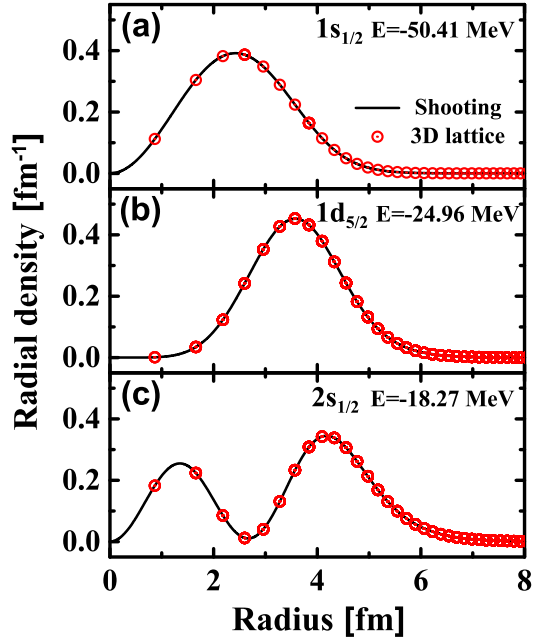}\\
  \caption{Radial density distributions for 1s$_{1/3}$, 1d$_{5/2}$, and 2s$_{1/2}$ in the 3D lattice calculation
  (open circles) in comparison with the shooting method (solid line). The radial density distribution in the 3D lattice calculation is extracted by Eq.\eqref{Eq_radialdensity}.
  }\label{Fig_radialdensity}
\end{figure}

To compare with the radial density distribution obtained by the shooting method, one can average the density distributions in the 3D lattice calculation,
\begin{equation}\label{Eq_radialdensity}
  \rho_{nlj}(r)=\frac{1}{2j+1} \sum_{i\in\{nlj\}}\psi_i^\dag(\bm{r})\psi_i(\bm{r}).
\end{equation}
In Fig.\ref{Fig_radialdensity}, the radial density distributions for 1s$_{1/3}$, 1d$_{5/2}$, and 2s$_{1/2}$ in the 3D lattice calculation (open circles) in comparison with the shooting method (solid line) are given,
in which a factor $4\pi r^2$ has been multiplied in order to amplify the radial density distribution at large distance. It can be clearly seen that the two distributions are in perfect agreement with each other.
The data points in the 3D lattice calculation are denser for large $r$ because the grid points used are uniform in the 3D lattice space.

\subsection{DEFORMED POTENTIALS}

\begin{figure}[h]
  \centering
  \includegraphics[width=0.45\textwidth]{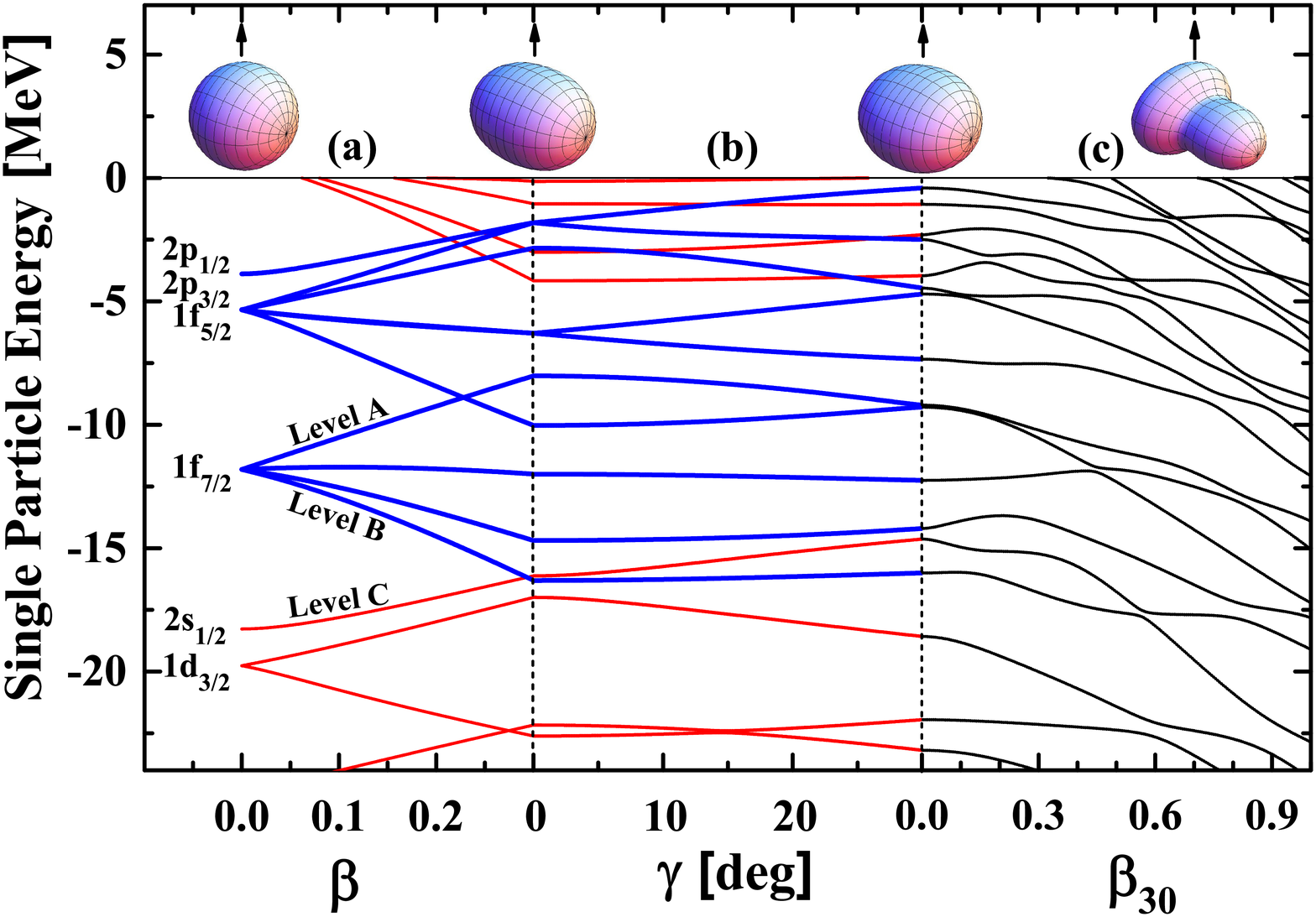}\\
  \caption{Single-particle levels in the deformed Woods-Saxon potential as functions of the deformation parameters $\beta$, $\gamma$, and $\beta_{30}$.
  The red and blue lines represent the levels with
  positive and negative parity respectively.
  The shapes shown in the top panel correspond to the deformed parameters
  $(\beta,\gamma,\beta_{30})=(0,0^\circ,0)$, $(0.3,0^\circ,0)$, $(0.3,30^\circ,0)$, $(0.3,30^\circ,0.7)$, respectively.
  }\label{Fig_deformedlevel}
\end{figure}

For the Dirac equations with the deformed potentials in Eq. \eqref{Eq_potential}, the single particle energies as functions of deformation parameters $\beta$, $\gamma$, and $\beta_{30}$ are given in Fig. \ref{Fig_deformedlevel}, which respectively correspond to axial, non-axial, and reflection-asymmetric deformed potentials.

In Fig. \ref{Fig_deformedlevel}(a), the potentials have both the
space reflection symmetry and axial symmetry with $\gamma=0$, $\beta_{30}=0$, and $\beta$ from $0$ to $0.3$.
In Fig. \ref{Fig_deformedlevel}(b), the potentials break the axial symmetry while keeping the space reflection symmetry
with $\beta=0.3$, $\beta_{30}=0$, and $\gamma$ from $0^\circ$ to $30^\circ$.
In Fig. \ref{Fig_deformedlevel}(c), the potentials break both the
space reflection symmetry and axial symmetry with $\beta=0.3$, $\gamma=30^\circ$, and $\beta_{30}$ from $0$ to $1.0$.

Although there is no symmetry restriction in the 3D lattice calculations, we can search for good quantum numbers from the expectations of physical operators.
For spherical cases, total angular momentum $j$ and orbital angular momentum $l$ can be calculated by the expectation of $\hat{\bm{j}}^2$ and $\hat{\bm{l}}^2$ with the upper components of the wave functions.
For axial cases, the $z$ component of the total angular momentum $|m_z|$ can be calculated by the expectation of  $\hat{j}_z^2$.
For the space reflection symmetry case, the parity can be calculated by the expectation of the parity operator $\hat{P}=\beta \hat{P}_{\bm{r}}$, where $\beta$ is the Dirac matrix and $\hat{P}_{\bm{r}} F(\bm{r})= F(\bm{-r})$.

From Fig. \ref{Fig_deformedlevel}, it can be seen that the levels in the spherical case are split into $(2j+1)/2$ levels with the potential changing from spherical to deformed. However, the Kramers degeneracy remains as there is no time odd potential.
For the axial case, the levels with lower (higher) $|m_z|$ values shift downwards (upwards) consistent with the
Nilsson model.
Comparing Fig. \ref{Fig_deformedlevel}(a) and Fig. \ref{Fig_deformedlevel}(b), it can be seen that the spectrum changes more modestly with $\gamma$ than with $\beta$.
In Fig. \ref{Fig_deformedlevel} (c), all levels trend to shift downwards with $\beta_{30}$, which shows its instability in fission.

\begin{figure}[h]
  \centering
  \includegraphics[width=0.45\textwidth]{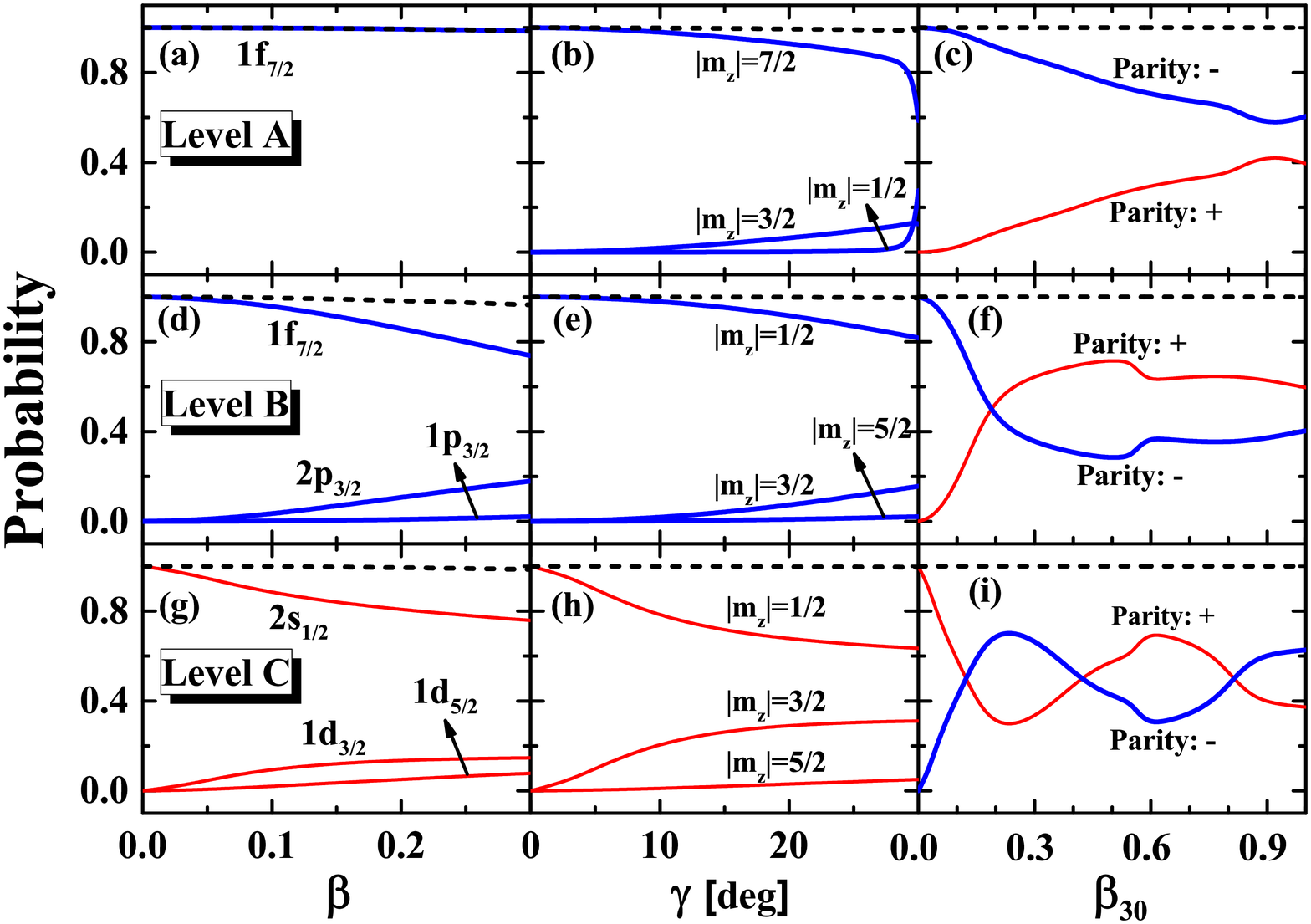}\\
  \caption{Compositions of levels A , B and C in Fig. \ref{Fig_deformedlevel} as functions of the deformation parameters $\beta$, $\gamma$, and $\beta_{30}$. The quantum numbers are given for each compositions. The total probabilities are shown as black dashed lines.}\label{Fig_components}
\end{figure}

To examine the compositions and their evolution of the single-particle level with deformation parameters $\beta$, $\gamma$, and $\beta_{30}$, levels A , B and C in Fig. \ref{Fig_deformedlevel} are chosen as examples.
 The results are illustrated in Fig. \ref{Fig_components}.
In the left panels, the compositions of each level are obtained by overlapping the wave functions with the wave functions obtained with $(\beta,\gamma,\beta_{30})=(0,0^\circ,0)$.
In the middle panels, the compositions of each level are obtained by overlapping the wave functions with the wave functions obtained with $(\beta,\gamma,\beta_{30})=(0.3,0^\circ,0)$.
In the right panels, the parity compositions of each level are obtained by the expectation of the parity operator.

In the left panels, there is only small mixing with other orbits for level A compared to levels B and C. It can be understood as follows. This is due to the special character of level A with $|m_z|=7/2$ and parity~$=-$.
The possible mixing is from the 1h$_{11/2}$ orbit which lies high in energy.
Similar conclusions can be drawn for the levels $|m_z|=3/2$ originating from 1p$_{3/2}$ and $|m_z|=5/2$ originating from 1d$_{5/2}$.

In the middle panels, for level A, there is a dramatic change for $|m_z|=7/2$ and $|m_z|=1/2$ components when $\gamma$ approaches $30^\circ$. This is due to the interaction between level A and the level originating from $1f_{5/2}$ and $|m_z|$=1/2 at $\gamma = 30^\circ$, as
shown in energy levels in Fig. \ref{Fig_deformedlevel}(b).

In the right panels, for the octupole deformed case, the parity composition of level B and C changes rigorously due to complicated interaction between levels.
For Level A, the main composition is negative-parity as it mainly interacts with negative-parity dominated levels.
All these can be understood from  Fig. \ref{Fig_deformedlevel} (c).

\section{SUMMARY AND PERSPECTIVES}\label{summary}

In summary, a new method to solve Dirac equation in 3D lattice space is proposed with the inverse Hamiltonian method to avoid variational collapse and the spectral method to avoid the Fermion doubling problem.
This method is demonstrated in solving the Dirac equation for a given spherical potential in 3D lattice space.
In comparison with the results obtained by the shooting method, the differences in single particle energies are smaller than $10^{-4}$~MeV, and the densities are almost identical, which demonstrates the high accuracy of the present method.
Applying this method to Dirac equations with an axial deformed, non-axial deformed, and octupole-deformed potential without further modification, the single-particle levels as functions of the deformation parameters $\beta$, $\gamma$, and $\beta_{30}$ are shown together with their compositions.

Efforts in implanting this method on the CDFT to investigate nuclei without any geometric restriction are in progress.
Possible applications include solving the Dirac equation in an external electric potential (deformation constrained calculation) to investigate nuclei with an arbitrary shape, and in an external magnetic potential (Coriolis term) to investigate rotating nuclei with arbitrary shape and an arbitrary rotating axis.
Moreover, the 3D time-dependent CDFT is also envisioned to be developed to investigate the relativistic effects in heavy-ions collisions and other nuclear reactions.

\begin{acknowledgments}
We thank P. Ring for helpful discussions.  This work was supported in part by the Major State 973 Program of China (Grant No. 2013CB834400), the National Natural Science Foundation of China (Grants No. 11335002, No. 11375015, No. 11461141002, No. 11621131001).
\end{acknowledgments}
%

\end{document}